# Pituitary Adenoma Segmentation

J. Egger, M. H. A. Bauer, D. Kuhnt, B. Freisleben and Ch. Nimsky

*Abstract*—Sellar tumors are approximately 10-15% among all intracranial neoplasms. The most common sellar lesion is the pituitary adenoma. Manual segmentation is a time-consuming process that can be shortened by using adequate algorithms. In this contribution, we present a segmentation method for pituitary adenoma. The method is based on an algorithm we developed recently in previous work where the novel segmentation scheme was successfully used for segmentation of glioblastoma multiforme and provided an average Dice Similarity Coefficient (DSC) of 77%. This scheme is used for automatic adenoma segmentation. In our experimental evaluation, neurosurgeons with strong experiences in the treatment of pituitary adenoma performed manual slice-by-slice segmentation of 10 magnetic resonance imaging (MRI) cases. Afterwards, the segmentations were compared with the segmentation results of the proposed method via the DSC. The average DSC for all data sets was 77.49%±4.52%. Compared with a manual segmentation that took, on the average, 3.91±0.54 minutes, the overall segmentation in our implementation required less than 4 seconds.

## I. Introduction

Approximately 10-15% of all intracranial neoplasms are sellar tumors. The most common sellar lesion is the pituitary adenoma [1]. The lesions can be classified according to size or hormone-secretion (hormone-active and horme-inactive). Microadenomas are less than 1 cm in diameter, whereas macroadenomas measure more than 1 cm. The rare giant-adenomas have more than 4 cm in diametric size. Secreted hormones can be cortisol (cushing's disease), human growth-hormone (hGH; acromegaly), follicle-stimulating hormone (FSH), luteinising hormone (LH), thyroid-stimulating hormone (TSH), prolactine or a combination of these. Only for the prolactine-expressing tumors, a pharmacological treatment is the initial treatment of choice in form of dopamine-agonists. Treatment is most commonly followed by a decrease of prolactine-levels and tumor volume. For acromegaly and cushing's disease, surgery remains the first-line treatment, although somatostatin receptor analogues or combined dopamine/somatostatin receptor analogues are a useful second-line therapeutical option for hGH-expressing tumors. Current medical therapies for Cushing's disease primarily focus on adrenal blockade of cortisol production, although pasireotide and cabergoline show promise as pituitary-directed medical therapy for Cushing's disease [2].

Thus, not only for the most hormone-active, but also for homone-inactive macroadenomas with mass-effect, surgery is the treatment of choice, most possibly via a transsphenoidal approach [3]. By contrast, for hormone-inactive mircroadenomas (<1cm) there is no need for immediate surgical resection. The follow-up contains endocrine and ophthalmological evaluation as well as magnetic resonance imaging (MRI). In case of continuous tumor volume progress, microsurgical excision becomes the treatment of choice. Thus, the tumor volume should be rigidly registered over the time of follow-up.

In this contribution, we present a segmentation method for pituitary adenoma. The method is based on an algorithm we developed recently in a previous work where the novel segmentation scheme was successfully used for segmentation of glioblastoma multiforme and provided an average Dice Similarity Coefficient (DSC) of 77%. For automatic and adequate adenoma segmentation, the original scheme is used, creating a directed 3D-graph within two steps: sending rays through the surface points of a polyhedron and sampling the graph's nodes along every ray. The center of the polyhedron is user-defined and located inside the adenoma. Then, the minimal cost closed set on the graph is computed via a polynomial time s-t-cut, creating an optimal segmentation of the adenoma's boundary and volume. For evaluation, neurosurgeons with strong experiences in the treatment of pituitary adenoma performed manual slice-by-slice segmentation of 10 cases. Afterwards, the segmentations were compared with the segmentation results of the proposed method via the Dice Similarity Coefficient (DSC). The average DSC for all data sets was 77.49%±4.52%. Compared with a manual segmentation that took, on the average, 3.91±0.54 minutes, the overall segmentation in our implementation required less than 4 seconds.

The paper is organized as follows. Section 3 presents the details of the proposed approach. In Section 4, experimental results are discussed. Section 5 concludes the paper and outlines areas for future work.

## II. Related Work

Based on magnetic resonance imaging, several algorithms have already been proposed for glioma segmentation. An extensive overview of some deterministic and statistical approaches is given by Angelini et al. [4]. The majority of them are region-based approaches; more recent ones are based on deformable models and include edge-information.

J.E., M.H.A.B., D.K., Ch.N. are with the Department of Neurosurgery, University of Marburg, Marburg, Germany e-mail: egger@med.uni-marburg.de
J.E., M.H.A.B., B.F. are with the Department of Mathematics and Computer Science, University of Marburg, Marburg, Germany

Neubauer et al. [5] and Wolfsberger et al. [6] introduce STEPS, a virtual endoscopy system designed to aid surgeons performing pituitary surgery. STEPS uses a semi-automatic segmentation method that is based on the so-called watershed-from-markers technique. The watershed-from-markers technique uses user-defined markers in the object of interest and the background. A memory efficient and fast implementation of the watershed-from-markers algorithm – also extended to 3D – has been developed by Felkel et al. [7].

Descoteau et al. [8] have proposed a novel multi-scale sheet enhancement measure and apply it to paranasal sinus bone segmentation. The measure has the essential properties to be incorporated in the computation of anatomical models for the simulation of pituitary surgery.

### III. METHODS

Our overall method starts by setting up a directed 3D graph from a user-defined seed point that is located inside the pituitary adenoma. To set up the graph, the method samples along rays that are sent through the surface points of a polyhedron with the seed point as the center. The sampled points are the nodes $n \in V$ of the graph $G(V,E)$ and $e \in E$ is the corresponding set of edges. There are edges between the nodes and edges that connect the nodes to a source $s$ and a sink $t$ to allow the computation of an s-t cut (the source and the sink $s, t \in V$ are virtual nodes).

The idea of setting up the graph with a polyhedron goes back to a catheter simulation algorithm where several polyhedra were used to align the catheter inside the vessel [9]. In the segmentation scheme, this idea is combined with a graph-based method that has been introduced for the semi-automatic segmentation of the aorta [10], [11], [12] and diffusion tensor imaging (DTI) fiber bundle segmentation [13]. However, in this case, setting up the graph was performed by sampling the nodes in several 2D planes and therefore is not useful for the segmentation of spherical or elliptical 3D objects (see Figure 1).

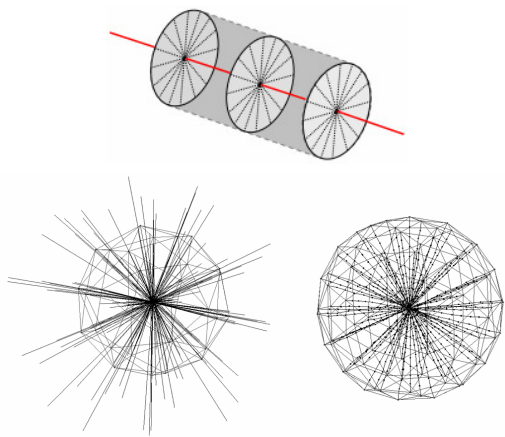

Fig. 1. Upper image: unfolding several 2D slices to segment a cylindrical object. Lower image: unfolding a polyhedron to segment a spherical object. Left: principal of sending rays through the surface of a polyhedron with 32 vertices. Right: Sampling nodes for the graph along the rays (polyhedron with 92 vertices).

In a recent publication [14], the introduced approach for segmenting spherical or elliptical 3D objects was enhanced by an arbitrary number of additional seed points to a semi-automatic method. Thereby, the algorithm is supported with grey value information and geometrical constraints that are used to segment World Health Organization (WHO) grade IV gliomas.

The arcs $<v_i, v_j> \in E$ of the graph $G$ connect two nodes $v_i$, $v_j$. There are two types of ∞-weighted arcs: z-arcs $A_z$ and r-arcs $A_r$ ($Z$ is the number of sampled points along one ray $z=(0,...,Z-1)$ and $R$ is the number of rays sent out to the surface points of a polyhedron $r=(0,...,R-1)$), where node $V(x_n,y_n,z_n)$ is one neighbor of node $V(x,y,z)$.

The arcs $A_z$ between two nodes along a ray ensure that all nodes below the polyhedron surface in the graph are included to form a closed set (correspondingly, the interior of the spherical object is separated from the exterior in the data). The arcs $A_r$ between the nodes of different rays constrain the set of possible segmentations and enforce smoothness via the parameter $\Delta_r$. The larger this parameter is, the larger the number of possible segmentations is (see Figure 2).

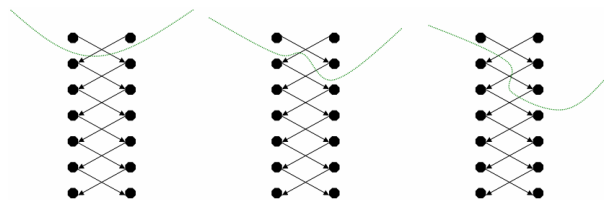

Fig. 2. Principle of a cut of edges between two rays for $\Delta_r=1$. Left and middle: Same cost for a cut (2·∞). Right: Higher cost for a cut (4·∞).

After graph construction, the minimal cost closed set on the graph is computed via a polynomial time s-t cut [15]. A Markov Random Field (MRF) approach where each voxel of the image is a node is definitely too time-consuming for the data we used (512x512xX). A MRF approach in a recent publication needed already several minutes for the cut of one small 2D image [16]. We also thought about using an Active Contour Method (ACM) [17] and [18] approach where the initial contour is a polyhedron with an initial radius definitely smaller than the object (pituitary adenoma). However, ACMs can get stuck in local minima during the iterative segmentation (expansion) process. In contrast, a graph cut approach provides an optimal segmentation for the constructed graph.

### IV. RESULTS

The presented methods were implemented in C++ within the MeVisLab platform [19]. Using 2432 and 7292 polyhedra surface points, the overall segmentation (sending rays, graph construction and mincut computation) in our implementation took less that 4 seconds on an Intel Core i5-750 CPU, 4x2.66 GHz, 8 GB RAM, Windows XP Professional x64 Version, Version 2003, Service Pack 2. The ray length is a fixed parameter (5 cm), determined via the largest pituitary adenoma of the 10 cases (all pituitary adenoma had a diameter of less that 5 cm).

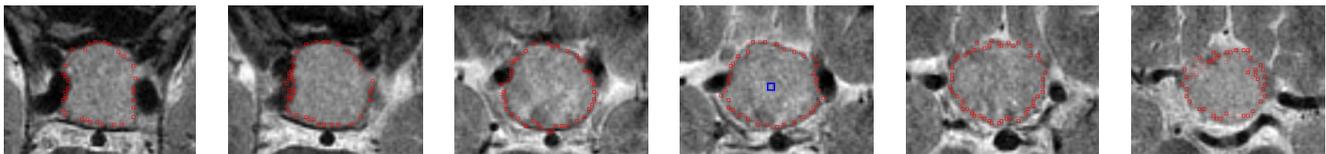

Fig. 3. Segmentation results for a pituitary adenoma data set with user-defined seed point (blue).

To evaluate the approach, neurological surgeons with several years of experience in resection of brain tumors performed manual slice-by-slice segmentation of 10 pituitary adenomas. Afterwards, the manual segmentations were compared with the one click segmentation results of the proposed method via the Dice Similarity Coefficient (DSC) [20]. The Dice Similarity Coefficient is the relative volume overlap between A and R, where A and R are the binary masks from the automatic (A) and the reference (R) segmentation. $V(\bullet)$ is the volume (in cm$^3$) of voxels inside the binary mask, by means of counting the number of voxels, then multiplying with the voxel size:

$$DSC = \frac{2 \cdot V(A \cap R)}{V(A) + V(R)} \quad (1)$$

In Figure 3, several axial slices of a pituitary adenoma are shown. In every slice, the result of the automatic segmentation is drawn (red). Additionally, the user-defined seed point is shown in the middle slice (blue). The average Dice Similarity Coefficient for all data sets was 77.49%±4.52% (see Table I for details). Figure 4 shows an axial slice of a pituitary adenoma (left) and a 3D mask of an automatic segmented pituitary adenoma (right). Figure 5 shows different views of sagittal slices with segmented pituitary adenoma performed with the proposed algorithm.

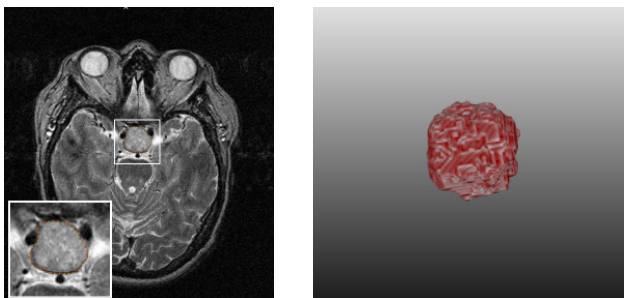

Fig. 4. Axial slice of a pituitary adenoma (left). 3D mask of an automatic segmented pituitary adenoma (right).

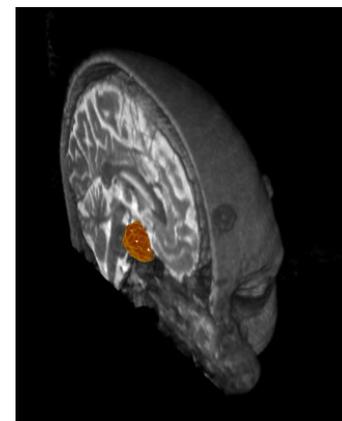

Fig. 5. Different views of sagittal slices with segmented pituitary adenoma.

## V. CONCLUSION

In this paper, a method for pituitary adenoma segmentation was presented. The method is based on an algorithm we developed recently in a previous work where the novel segmentation scheme was successfully used for segmenting glioblastoma multiforme and provided an average Dice Similarity Coefficient (DSC) of 77%. For automatic and adequate adenoma segmentation, the original scheme was used, creating a directed 3D-graph within two steps: sending rays through the surface points of a polyhedron and sampling the graph's nodes along every ray. The center of the polyhedron is user-defined and located inside the adenoma. Then, the minimal cost closed set on the graph is computed via a polynomial time s-t-cut, creating an optimal segmentation of the adenoma's boundary and volume.

For evaluation, neurosurgeons with strong experiences in the treatment of pituitary adenoma performed manual slice-by-slice segmentation of 10 cases. Afterwards, the segmentations were compared with the segmentation results of the proposed method via the Dice Similarity Coefficient. The average Dice Similarity Coefficient for all data sets was 77.49%±4.52%. Compared with a manual segmentation that took, on the average, 3.91±0.54 minutes, the overall segmentation in our implementation needed less than 4 seconds.

TABLE I
SUMMARY OF RESULTS: MIN., MAX., MEAN AND STANDARD DEVIATION FOR 10 PITUITARY ADENOMAS.

|  | Volume of tumor (cm$^3$) | | Number of voxels | | DSC (%) | manual segmentation time (min) |
|---|---|---|---|---|---|---|
|  | manual | algorithm | manual | algorithm |  |  |
| min | 0.84 | 1.18 | 4492 | 3461 | 71.07 | 3 |
| max | 15.57 | 14.94 | 106151 | 101902 | 84.67 | 5 |
| $\mu \pm \sigma$ | 6.30 ± 4.07 | 6.22 ± 4.08 | 47462.7 | 47700.6 | 77.49 ± 4.52 | 3.91 ± 0.54 |

There are several areas of future work. For example, we want to enhance the presented segmentation scheme with statistical information about shape – e.g. Active Shape Models (ASM) [21] – ort texture – e.g. Active Blobs [22] – or shape and texture – e.g. Active Appearance Models [23] and [24] – of the desired object. Furthermore, we want use the approach to segment spherically or elliptically shaped organs like kidneys [25] and pathologies like cerebral aneurysms [26] and [27].

## VI. Acknowledgements

The authors would like to thank Fraunhofer MeVis in Bremen, Germany, for their collaboration and especially Horst K. Hahn for his support.